# Nanoindentation of Porous Bulk and Thin Films of La$_{0.6}$Sr$_{0.4}$Co$_{0.2}$Fe$_{0.8}$O$_{3-\delta}$

Zhangwei Chen*, Xin Wang, Vineet Bhakhri, Finn Giuliani, Alan Atkinson
Department of Materials, Imperial College London, London SW7 2AZ, United Kingdom
Corresponding author: chen@ic.ac.uk

## Abstract

In this paper we show how reliable measurements on porous ceramic films can be made by appropriate nanoindentation experiments and analysis. Room-temperature mechanical properties of the mixed-conducting perovskite material La$_{0.6}$Sr$_{0.4}$Co$_{0.2}$Fe$_{0.8}$O$_{3-\delta}$ (LSCF6428) were investigated by nanoindentation of porous bulk samples and porous films sintered at temperatures from 900-1200 °C. A spherical indenter was used so that the contact area was much greater than the scale of the porous microstructure. The elastic modulus of the bulk samples was found to increase from 33.8-174.3 GPa and hardness from 0.64-5.32 GPa as the porosity decreased from 45-5% after sintering at 900-1200 °C. Densification under the indenter was found to have little influence on the measured elastic modulus. The residual porosity in the "dense" sample was found to account for the discrepancy between the elastic moduli measured by indentation and by impulse excitation. Crack-free LSCF6428 films of acceptable surface roughness for indentation were also prepared by sintering at 900-1200 °C. Reliable measurements of the true properties of the films were obtained by data extrapolation provided that the ratio of indentation depth to film thickness was in the range 0.1 to 0.2. The elastic moduli of the films and bulk materials were approximately equal for a given porosity. The 3D microstructures of films before and after indentation were characterized using FIB/SEM tomography. Finite element modelling of the elastic deformation of the actual microstructures showed excellent agreement with the nanoindentation results.

Key words: LSCF cathode films; microstructure; nanoindentation; mechanical properties; FEM.

## 1. Introduction

Over the last two decades, lanthanum-based perovskite-structured materials with general formula La$_{1-x}$Sr$_x$Co$_y$Fe$_{1-y}$O$_{3-\delta}$, denoted as LSCF, have been extensively studied due to their promising mixed electronic-ionic conductivity (MEIC)[1] and high oxygen surface exchange rate[2] for applications in cathodes of intermediate-temperature solid oxide fuel cells (IT-SOFCs)[3] and for oxygen separation membranes[4]. These functional components must withstand mechanical stresses during fabrication and operation in order to achieve long-term durability and reliability. Therefore, suitable mechanical properties are desired to prevent failures such as cracks, delamination and fractures due to mechanical stresses arising from thermal expansion coefficient differences, temperature gradients, oxygen pressure gradients and external mechanical loading [5, 6]. Such damage even if not mechanically catastrophic, usually results in degradation of electrochemical performance. Therefore, it is imperative to understand the mechanical stability of these materials in the form in which they are deployed in applications which, in the case of a fuel cell or electro-catalyst, is as a porous film on a dense substrate. One of the key mechanical properties is the elastic modulus and this is the subject of this study.

To date, most of the studies on LSCF have concentrated on electrical properties[1, 7], oxygen permeability, diffusion and transport[8-10], degradation mechanisms[11, 12], thin film synthesis[13, 14] and applications[15, 16]. There are only a few reports of their mechanical properties. The room temperature Young's and shear moduli, hardness, fracture toughness and biaxial flexure strength of nominally dense, bulk $La_{1-x}Sr_xCo_{0.2}Fe_{0.8}O_3$ (x=0.2-0.8) were reported by Chou et al.[17]. Young's moduli of 151-188 GPa were determined using the impulse excitation method. They found that increasing Sr content increased the Young's and shear moduli. Young's modulus and strain-stress behaviour of nominally dense bulk $La_{0.5}Sr_{0.5}Fe_{1-y}Co_yO_{3-\delta}$ ($0 \leqslant y \leqslant 1$) in the temperature range of 20-1000 °C have been studied by Lein et al.[18]. They report room-temperature Young's modulus measured by impulse excitation to be 130±1 GPa, for y = 0.2. A nonlinear stress–strain relationship was observed in four-point bending experiments at room temperature and was inferred as a signature of ferroelastic behaviour. Above the ferroelastic to paraelastic transition temperature (~900 °C), the materials showed linear elastic behaviour. The ferroelasticity of perovskite materials had been reported earlier by Kleveland and co-workers[19, 20] and Faaland et al.[21]. Huang et al. [22] measured the mechanical properties of $La_{0.58}Sr_{0.4}Co_{0.2}Fe_{0.8}O_{3-\delta}$ using ring-on-ring biaxial flexure and found that the measured room temperature fracture stress was nearly 40% higher than that at 800°C. They ascribed this to the ferroelasticity of the material at room temperature. However, the literature data above are based on nominally dense bulk materials, which are quite different from the highly porous films used in most applications. Moreover, most studies have employed conventional macroscopic techniques to measure the mechanical properties, such as resonance methods, bending tests, compression and tensile tests, but these techniques are not applicable for thin films coated substrates.

The nanoindentation technique has been developed and used extensively to measure the mechanical properties of small volumes of material, including thin films, due to the potentially high spatial and depth resolution of the measurement[23, 24]. Despite this, there is currently no study available on nanoindentation to characterise the micromechanical properties of LSCF, neither in the form of bulk samples nor as porous films. Porous films may behave mechanically very differently from bulk samples of the same materials/compositions, and therefore a method to obtain the film-only properties is desirable. However, the methodology for extracting "long range" mechanical properties from indentation of porous films has not been properly established. Previous studies on nanoindentation of porous ceramic films[25, 26], polymeric coatings[27-29] and highly porous bioceramic layers such as bones[30-32] have adopted simple rules-of-thumb, such as indenting to less than 10% of the film thickness. Furthermore, no significant work has been done on characterising elastic modulus and/or hardness as a function of porosity in partially sintered ceramics/films. More recently, studies of porous films for microelectronic applications using nanoindentation with Berkovich indenters have been reported [33, 34]. However, their methodology is not suitable for typical partially sintered ceramic films. This is because their films were very smooth, flat and ductile polymers having low porosity (< 30 vol %) and extremely small pores in the nm range. Therefore a sharp Berkovich indenter was able to deform a representative volume of the porous film. However, partially sintered ceramic films (such as the ones in the current study) often feature a relatively rough surface and have pores in the micron range and porosity as high as 50 vol%. Furthermore they are not ductile and have a low elastic limit. The scale of their microstructures requires use of a blunt indenter in order to deform a representative volume of material in the film and avoid the sensitivity of sharp indenters to surface roughness. The brittleness also implies a densification mechanism by crushing, which is significantly different from the ductile deformation of polymer films.

In the current paper we describe experiments on highly porous ceramic films and bulk specimens using a spherical indenter. For this purpose LSCF6428 powders were sintered at temperatures ranging from 900-1200 °C in both bulk and thin film forms to obtain varying microstructures. We then present a methodology to extract reliable values for the elastic modulus and hardness of the films and verify the elastic moduli by FEM of 3D reconstructed microstructures obtained using FIB/SEM tomography.

## 2. Materials and Methods

### 2.1 Sample preparation

*2.1.1 LSCF6428 and $Ce_{0.9}Gd_{0.1}O_{2-\delta}$ (CGO) pellets:*

Dense LSCF6428 and CGO pellet samples were prepared for mechanical property characterisation and as substrates for film deposition, respectively, by the following steps. LSCF6428 and CGO powders obtained from NexTech Materials, USA, were first uniaxially pressed separately into disc shapes by applying 150 MPa load. They were then isostatically pressed at 200 MPa for further material compaction, followed by sintering (at 900, 1000, 1100 and 1200 °C for LSCF6428, and 1400 °C for CGO) in air with a heating rate of 5 °C/min and a holding time of 4 hours. The as-sintered bulk samples were finally polished on one face by successively using 15, 9, 6, 4 and 1 micron diamond suspensions, in order to generate mirror smooth and flat surfaces.

*2.1.2 LSCF6428 inks and films*:

A commercial LSCF6428 screen-printing ink provided by ESL-UK was diluted 1:2 by volume with terpineol (Sigma, UK) and then ball-milled to give a homogeneous slurry suitable for tape casting. Both the original commercial ink and the reformulated (diluted) ink were used to fabricate films. The films were made by tape casting the inks onto the polished surface of the CGO pellets using a perimeter mask of 40 μm height. The films were dried for 12 hours at 100 °C and then sintered at 900, 1000, 1100 or 1200 °C in air for 4 hours with a heating rate of 5 °C/min.

## 2.2 Sample characterisation

*2.2.1 Relative density, porosity and surface roughness measurement:*

The relative densities as well as porosities of as-sintered LSCF6428 bulk samples were measured using an Archimedes' balance. The porosities of LSCF6428 films after sintering at 900 - 1200 °C were measured using the Statistics module of Avizo 6.0 image processing software (VSG Co., USA) based on the actual 3D microstructural data of the films collected using FIB/SEM technique. The surface roughness of the as-sintered LSCF6428 films was measured using an optical interference surface profiler (Zygo, USA).

*2.2.2 Elastic modulus and hardness measurement using nanoindentation:*

The nanoindentation experiments on the as-sintered films and bulk samples were performed on a NanoTest platform (Micromaterials, UK) at room temperature. Both a Berkovich diamond tip and a spherical diamond tip of 50 μm diameter were used in this study depending on the sample forms under investigation. Note that compared to sharp indenters including Berkovich tips, spherical tips facilitate the distinction from elastic to plastic deformation of materials during indentation due to their less drastic variation of stress under loading. The spherical tip is particularly appropriate for porous materials as the deformation zone can be arranged to be of much greater length scale than the typical length scale of the porous microstructure (e.g. the average pore diameter). They thus can give a result that characterises the long range properties of the porous material. Other benefits of using spherical tips include less sensitivity to surface condition and more accurate resultant hardness [35]. At least 20 measurements were conducted in different locations for each sample in order to measure the variability of the mechanical response of the sample. Prior to nanoindentation tests, the NanoTest platform was precisely calibrated using a standard silica sample to establish the system frame compliance.

According to the Oliver-Pharr method[23], the hardness $H$ and elastic modulus $E_0$ of the sample are given by the following equations based on the load ($P$) versus indentation depth ($h$) test curve:

$$H = \frac{P_{max}}{A} \tag{1}$$

$$E_0 = \frac{1-\nu_0^2}{\dfrac{2}{\beta S}\sqrt{\dfrac{A}{\pi}} - \dfrac{1-\nu_i^2}{E_i}} \tag{2}$$

where $P_{max}$ is the peak load, $A$ is the projected contact area, $S$ is the stiffness representing the slope of the initial unloading part of the curve, $\nu$ is the Poisson's ratio, and $\beta$ is a factor that depends on the indenter shape ($\beta = 1$ for a spherical indenter and 1.034 for a Berkovich one). The terms shown with subscript $0$ and $i$ are the properties of the test sample and indenter, respectively. A sensitivity study showed that the elastic modulus calculated from indentation depends little (e.g. variation < 8%) on the variation of Poisson's ratio from 0.2-0.4[36]. As a result, $\nu_0 = 0.31$ of the fully dense LSCF6428 material reported in [17] was used in Equation (2) for all specimens.

*2.2.3 Elastic modulus measurement of dense LSCF6428 samples by impulse excitation:*

For comparison, the elastic modulus of the dense LSCF6428 pellets after sintering at 1200 °C was also measured using the impulse excitation technique (IET), which is a dynamic and macroscopic method as opposed to nanoindentation. The measurements were conducted on polished LSCF6428 dense pellets which were 25 mm in diameter and approximately 2 mm thick, using a GrindoSonic MK5 resonance system (J.W. Lemmens, Belgium), according to British Standard EN 843-2:2006[37]. This involves measuring two natural vibrational mode frequencies of the sample, $f_1$ and $f_2$, from which two dynamic Young's moduli, $E_1$, $E_2$, were calculated using the following equation [37]:

$$E_{i,i=1,2} = 12\pi f_i d_i^2 m_i^2 \frac{1-\upsilon^2}{K_i^2 t_i^3} \qquad (3)$$

where $d$ = the pellet diameter; $t$ = the pellet thickness; $m$ = the pellet mass; $f$ = frequency; $K$ = the geometric factor of the vibration mode; $v$ = the Poisson's ratio. The Poisson's ratio along with values of $K$ were determined according to the parameter tables from [37]. The average value of dynamic Young's modulus, $E$, was determined by taking the average of $E_1$ and $E_2$.

*2.2.4 Microstructural characterisation:*

The microstructures of the samples and thickness of the films were characterised by scanning electron microscopy (JSM-5610LV SEM, JEOL, Japan). The average grain sizes of the samples were measured based on analysis of the micrographs obtained. A focused ion beam/scanning electron microscope (FIB/SEM) dual beam instrument (Helios NanoLab 600i, FEI, USA) was employed in this study to perform 3D microstructural characterisation of the as-sintered and as-indented samples. Prior to the FIB/SEM characterisation, the films were vacuum-impregnated with low-viscosity epoxy resin in order to obtain high contrast images on well-defined planar surfaces (i.e. avoiding penetration of the electron beam into the pores) and at the same time to protect the interconnected porous structures of the films from damage.

*2.2.5 Finite element modelling (FEM):*

FEM was performed using Abaqus CAE 6.10 (Dassault Systemes, USA) to calculate the effective elastic modulus of the actual 3D microstructures of the as-sintered films. The 3D microstructures were obtained by 3D reconstructions using Avizo 6.0 based on the stacks of sequential 2D images recorded by SEM of the FIB-sliced cross-sections of the films. The distance between two adjacent images was set to be 50 nm. The reconstruction involved binary segmentation and alignment of the image stacks, the resulting 3D data of which were then imported into ScanIP package (Simpleware, UK) for smooth hexahedral and tetrahedral meshing in order to generate 3D finite element models. The FEM was run for each microstructure (corresponding to a given sintering temperature) using the Abaqus Standard FE solver based on the assumption that LSCF6428 solid material was isotropic, linear and elastic, irrespective of the sintering temperature. A small displacement or constrained boundary condition was applied as required to the model surfaces. After simulation, the average normal force on the displaced surface area (solid plus porosity) was obtained to calculate the effective elastic modulus of the 3D microstructure. It is worth noting that the FEM modelled the real microstructures based on 3D reconstruction, rather than an equivalent continuous medium model used in [33, 38].

# 3. Results and Discussion

## 3.1 Crack elimination and surface quality improvement for LSCF6428 films

In the current study, extensive cracks and considerable surface roughness were generated in the as-sintered films made from the commercial LSCF6428 ink. Examples are shown in Figures 1 (a) and (b) for a film which was sintered at 1000 °C. Such cracking is not necessarily a problem in some applications, but it was critical for the present study. It not only made the films vulnerable to further damage during handling, but significantly affected the nanoindentation response and was unfavorable for the FIB/SEM tomography and the subsequent FEM process. Moreover, the poor smoothness of the film surface gave an unreliable initial contact leading to considerable scatter in the nanoindentation data (e.g. relative error > 30% for the film displayed in Figures 1 (a) - (b)).

Our recent study on crack formation in fuel cell electrode films[39] has revealed that rather than shrinkage during sintering, or differential contraction during cooling as reported in many studies[40, 41], the critical factor for obtaining crack-free and smooth films in this study was the ability of the ink to be self-leveling in the earlier wet state. Cracking was initiated at the drying stage if the particles were prevented from packing more effectively as the liquid was removed. As a result, by lowering the viscosity of the ink, cracking can be reduced to an acceptable level or even be completely avoided. In the present study this was achieved using reformulated ink by diluting as described in Section 2. Crack-free films with smooth surface were successfully fabricated after sintering at 1000 ºC using reformulated ink, as shown in Figures 1 (c) and (d), which led to a remarkable drop of surface roughness $R_a$ from 1.80 μm to 0.20 μm. As a result, reproducible measurements were ensured with much less scatter (relative error < 8%) in the data compared to the films of poor surface quality described earlier.

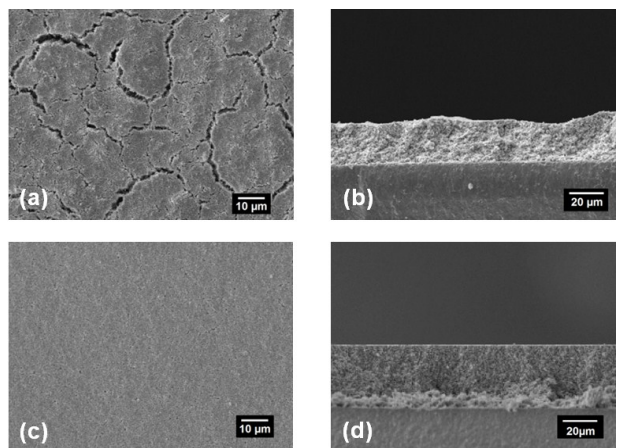

*Fig.1. Micrographs of top surface and cross-section of films made using (a-b): original commercial ink; (c-d): reformulated ink.*

### 3.2 Porosity and SEM microstructures of LSCF6428 films and bulk samples

The porosities measured for LSCF6428 in both film and bulk forms sintered at 900-1200 °C are summarized in Table 1. It is found that sintering at 900 °C tended to generate approximately 45 % porosity for both forms of samples. After sintering at 1000 and 1100°C, both the films and bulk samples had similar porosities. However, the bulk porosity experienced a huge drop to only 5 % after sintering at 1200 °C, compared to 15 % for films. These trends in the evolution of porosity as a function of sintering temperature can be readily seen in the micrographs shown later.

*Table 1. Porosity vs. sintering temperature for LSCF6428 films and bulk samples.*

| Sintering Temperature (°C) | Film Porosity (%) | Film average grain size (nm) | Bulk Porosity (%) |
|---|---|---|---|
| 900 | 46.9±2.2 | 200 | 44.85±0.32 |
| 1000 | 39.7±2.6 | 270 | 36.28±1.12 |
| 1100 | 24.1±1.8 | 450 | 28.67±0.95 |
| 1200 | 15.2±1.2 | 690 | 5.22±0.01 |

The surface features of the as-sintered films are shown in Fig. 2. The average grain sizes of these films were measured and are shown in Table 1, indicating that higher sintering temperatures resulted in coarser grains.

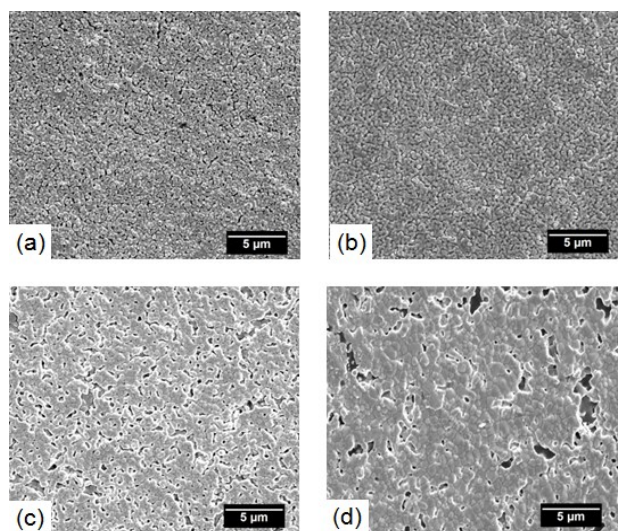

*Fig.2. Micrographs of top surface of LSCF6428 films after sintering at (a) 900; (b) 1000; (c) 1100; and (d) 1200 °C.*

As can be seen in Fig. 2, little densification or grain growth took place during sintering at 900 and 1000 °C, consistent with the large porosity shown in Table 1. The increasingly large pores observed on the surface with higher sintering temperature are typical of films formed by constrained sintering [39]. It is also apparent that there are no detectable surface cracks in these films for all sintering temperatures. The evolution of the corresponding cross-sectional microstructures shown in Fig.3 is consistent with the images from the top surfaces.

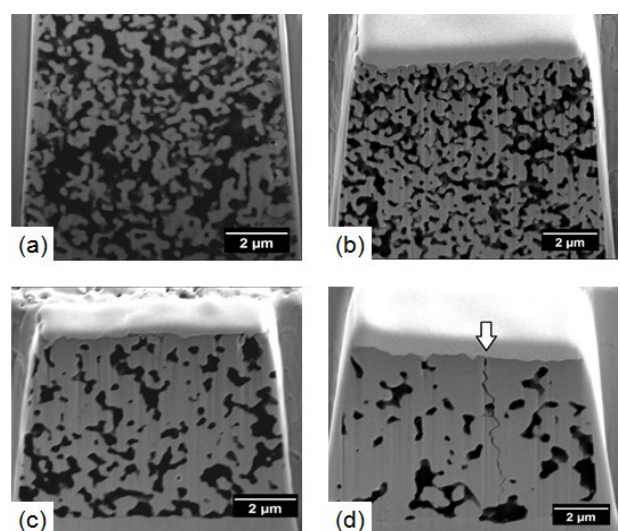

*Fig.3. FIB-milled cross-sectional micrographs of LSCF6428 films after sintering at (a) 900; (b) 1000; (c) 1100; and (d) 1200 °C (black represents porosity and gray is LSCF6428).*

A very narrow microcrack penetrating through the film can be seen in Fig. 3 (d) (marked with an arrow). The narrowness of the crack indicates that it was formed after sintering, probably due to thermal contraction mismatch on cooling. Such a crack might cause errors if the nanoindentation test was conducted nearby, but these cracks are rare and any individual indentations affected by them would be apparent in the distribution of measured values. In the same way, the SEM micrographs of the polished top surface and FIB-milled cross-sections of each as-sintered bulk sample are shown in Figures 4 and 5. Note that resin impregnation was not performed for these samples.

These microstructures are consistent with the porosity measurements shown in Table 1. The porosity-dependent mechanical properties of both film and bulk samples, which are the main concern of this paper, will be described later on.

### 3.3 Elastic modulus and hardness of LSCF6428 bulk samples measured using nanoindentation

The elastic modulus and hardness of bulk samples measured using nanoindentaion are shown in Figures 6 and 7 plotted against maximum indent load, which is approximately proportional to the maximum indentation depth.

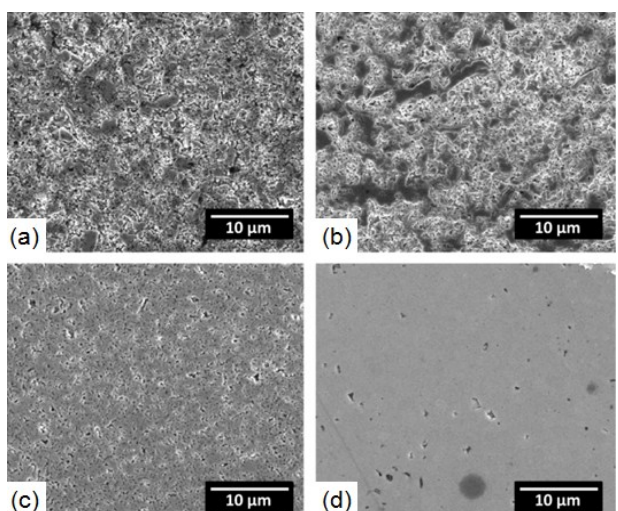

*Fig.4. Polished surface SEM micrographs of bulk samples after sintering at (a) 900; (b) 1000; (c) 1100; and (d) 1200 °C.*

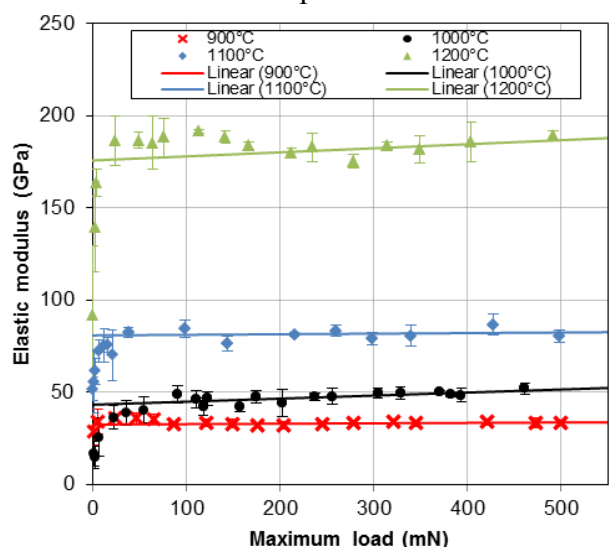

*Fig.6. Elastic modulus vs. maximum indent load for bulk samples after sintering at 900 - 1200 °C. The solid lines represent the extrapolations to zero load.*

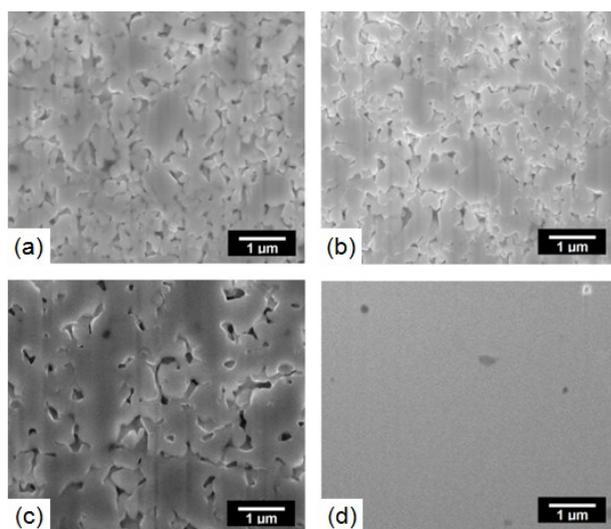

*Fig.5. FIB-milled cross-sectional micrographs of bulk samples after sintering at (a) 900; (b) 1000; (c) 1100; and (d) 1200 °C.*

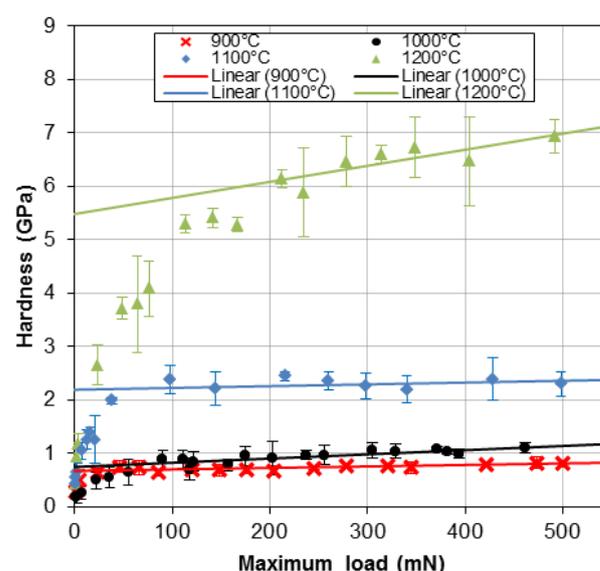

*Fig.7. Hardness vs. maximum indent load for bulk samples after sintering at 900 - 1200 °C. The solid lines represent the extrapolations to zero load.*

Fig. 6 clearly shows that, in all the bulk samples, the apparent elastic modulus exhibits an initial dramatic increase with increasing maximum indent load below approximately 50 mN, followed by a levelling off at higher loads, except the existence of turbulence between 50-200 mN for 1200 °C data. The increasing elastic modulus at < 50 mN was thought to be attributed to the surface roughness being comparable to the indentation depth at low loads. Surface roughness could induce underestimation of $E$ and $H$ at shallow indentation depth (i.e. small load), as shown in Figures 6 and 7. This underestimation was due to the overestimated contact area, provided that the indentation depth was comparable to the roughness, as also reported by other investigations[42, 43]. The measured $E$ and $H$ increased with the indentation depth as the surface roughness effect gradually diminished when an increasing number of the untouched valleys close to the asperities on the surface under compression were touched by the indenter tip. Such effect can be eliminated in data analysis by simply discarding the indentation depths (or loads) smaller than a critical value, for which the result are less reliable, as suggested by Mencik and Swain[44]. Therefore in our case, the best estimate of true elastic modulus was obtained by extrapolating the data points from the load range 50 – 500 mN to zero load for 900-1100 °C data and 200-500 mN for 1200 °C data, as shown by the solid lines in Fig. 6. The same method was used to obtain the best estimates of the true hardness. The estimated elastic modulus and hardness are summarised in Table 2.

The observation that for loads > 50 mN there is only a weak dependence of apparent modulus on load, indicates that densification (crushing) of the porous bulk material under the indenter had a negligible influence on the elastic response.

*Table 2. Elastic moduli and hardness of the as-sintered bulk samples.*

| Sintering Temperature (°C) | Elastic modulus (GPa) | Hardness (GPa) |
|---|---|---|
| 900 | 34.2±2.1 | 0.69±0.09 |
| 1000 | 44.5±3.2 | 0.86±0.20 |
| 1100 | 80.2±1.9 | 2.35±0.14 |
| 1200 | 174.3±2.8 | 5.76±0.12 |

In principle, densification of the material generated by crushing under indenter could significantly increase the local elastic modulus. This lack of sensitivity to densification is presumably because the densified zone is small compared with the longer range of elastic deformation of the rest of the material. This phenomenon will be examined in more detail later in the data analysis of the porous films, for which the factors influencing the measurements are more complicated. However, it can be seen from Fig. 7 that the hardness is more sensitive to indenter load, implying that the hardness was more influenced by densification. This is to be expected since the hardness (plastic deformation) of these materials is controlled by crushing rather than the more usual plastic deformation (viscous flow or dislocation motion) seen in other materials.

### 3.4 Comparison of IET and indentation results for dense LSCF6428 samples

The elastic modulus of "dense" LSCF6428 samples was also measured by IET to compare with the nanoindentation result after the accuracy of the IET measurements was calibrated. The elastic modulus measured by IET for the "dense" pellet was 147±3 GPa, which is close to, but slight

*Table 3. Comparison of Young's modulus measurements for "dense" LSCF6428 samples.*

| Reference | Sintering Conditions (in air) | Relative Density (%) | Main Grain Size (µm) | Measurement Technique | Young's Modulus at RT (GPa) | Hardness (GPa) |
|---|---|---|---|---|---|---|
| Kimura et al. [45] | 1300°C/6h/106°C·h⁻¹ | 98 | 5.0 | IET | 164 | n/a |
| Chou et al. [17] | 1250°C/4h/300°C·h⁻¹ | 95.4±0.2 | 2.9 | Ultrasonic method | 152±3 | n/a |
| This work | 1200°C/4h/300°C·h⁻¹ | 94.78±0.01 | 1.6 | IET | 147±3 | n/a |
| | | | | Nanoindentation (Berkovich tip, $P_{max}$=500mN) | 180±10 | 7.0±0.2 |
| | | | | Nanoindentation (Spherical tip, $R$=25µm, $P_{max}$=500mN) | 174±3 | 5.3±0.1 |

-ly lower than the results reported by Chou et al.[17] (152±3 GPa) and Kimura et al.[45] (164 GPa), as shown in Table 3.

This might be due to differences in relative densities or chemical composition between the samples used. In the present nanoindentation experiments, using a spherical tip tended to generate slightly smaller values of both elastic modulus and hardness, compared to the use of the Berkovich tip. This is expected since indentation modulus with a non-axisymmetric sharp indenter is typically a few percent larger than that obtained using axisymmetric indenters, as revealed by Vlassak and Nix [46]. In addition, the resulted standard deviations also confirm that compared with the Berkovich tip, the spherical tip was more consistent, which might be due to it being less sensitive to the sample surface condition. Nevertheless, it is clear that the indentation method gave significantly higher modulus than the other methods and we have considered whether non-elastic behaviour might be responsible for this.

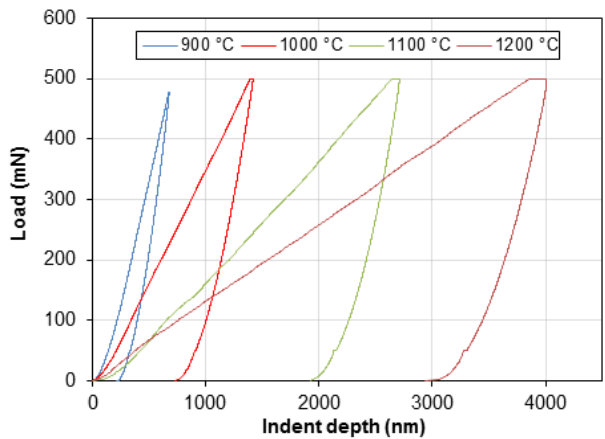

*Fig. 8. Smooth load-depth curves shows absence of pop-in or pop-out events in the nanoindentation response for bulk samples sintered at varying temperatures.*

LSCF6428 has a rhombohedral perovskite structure at room temperature[3, 47], with a transition to cubic perovskite occurring at approximately 773 K[47]. Therefore, any phase transition of the material due to temperature difference can be ruled out as a source of the discrepancy, as the IET and nanoindentation tests were performed at room temperature. We therefore considered other possibilities such as: (1) ferroelastic behaviour induced by the indentation stresses; (2) micro/macroscopic structural defects, including microcracks and porosity in the sample.

However, no evidence of non-elastic pop-in and pop-out or hysteresis was found during indentation even when the load applied was larger than 500 mN. Fig. 8 shows a set of typical load vs. depth curves suggesting very smooth nanoindentation response for bulk samples sintered at varying temperatures.

The reason for ferroelasticity not being observed during indentation may be due to the domain dimension being limited by the small grain size thereby pinning domain walls. Further study is required to clarify this. Furthermore, no evidence of microcracking in the present specimens used in IET experiments could be detected by SEM. Nevertheless, there was approximately 5% of open porosity in the "dense" LSCF6428 samples. The effect this might have on the elastic modulus was estimated by applying the model proposed by Ramakrishnan and Arunachalam [48] (composite sphere method) for solids with randomly distributed isolated pores, namely:

$$E_0 = \frac{E_p(1 + b_E p)}{(1 - p)^2} \qquad (4)$$

In Eq.(4), $E_p$ is the modulus for porous material having porosity, $p$, and $E_0$ is the modulus of the fully dense material. $b_E$ is a parameter depending on Poisson's ratio $v_0$ of fully dense material, with $b_E = 2 - 3v_0$. The values of $E_p$ measured by the IET method and the corresponding porosities were extrapolated to fully dense material using Eq.(4) and the results are shown in Table 4.

*Table 4. Extrapolated elastic moduli ($E_0$) of "dense" LSCF6428 obtained by applying Eq.(4).*

| Reference | Porosity (%) | Measured Elastic Modulus $E_p$ (GPa) | Extrapolated Elastic Modulus $E_0$ (GPa) |
|---|---|---|---|
| Kimura et al. [45] | 2 | 164 | 175 |
| Chou et al. [17] | 4.64±0.22 | 152±3 | 176±3 |
| This work | 5.22±0.01 | 147±3 | 173±3 |

The extrapolated values for fully dense material are all in good agreement within experimental error. Furthermore, the extrapolated results are also close to the indentation results. Therefore we can conclude that the reason for the discrepancy between moduli measured by IET and by indentation is residual porosity in the supposedly "dense" bulk specimens. The IET method samples the entire specimen volume to generate the appropriate "long-range" modulus for the material and is affected by residual porosity, whereas the indentation method only samples locally fully dense regions, resulting in the true modulus. This was recognized by some other studies [49, 50], which reported that the existence of porosity was responsible for the large discrepancies between elastic moduli of polycrystalline solids measured using "bulk" methods (such as IET) and nanoindentation.

### 3.5 Elastic modulus and hardness of porous LSCF6428 films measured using nanoindentation

Figures 9 and 10 show the nanoindentation data for elastic modulus and hardness of the films as a function of the ratio of maximum indentation depth relative to film thickness (i.e. $h_{max}/t_f$).

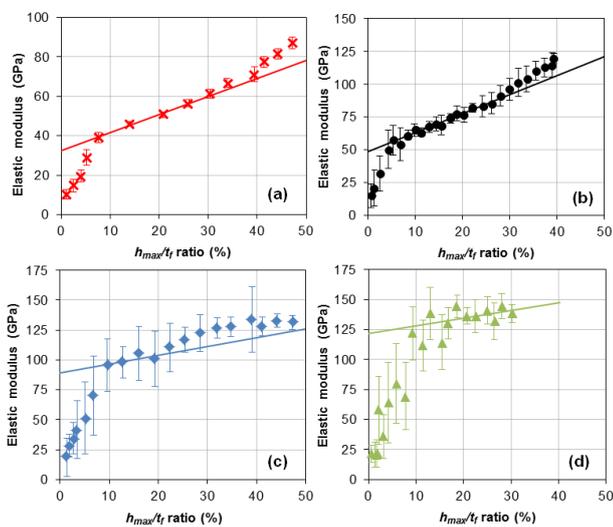

Fig.9. *Elastic modulus vs. $h_{max}/t_f$ ratio for porous films after sintering at: a) 900; b) 1000; c) 1100; and d) 1200 °C. The solid lines represent the extrapolations to zero depth (i.e. zero load).*

The results show a clear dependence of elastic modulus and hardness on indentation depth, which reflect the combined effects from surface roughness, densification and the substrate. This situation is clearly more complicated than for the bulk samples discussed earlier.

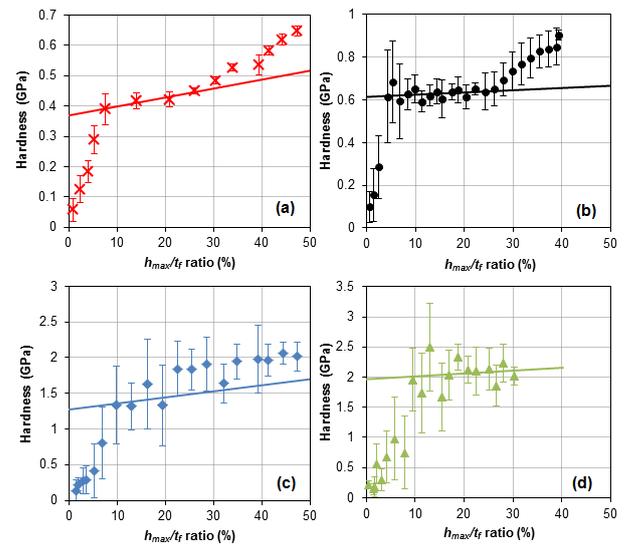

Fig.10. *Hardness vs. $h_{max}/t_f$ ratio for porous films after sintering at: a) 900; b) 1000; c) 1100; and d)1200 °C The solid lines represent the extrapolations to zero depth (i.e. zero load).*

#### 3.5.1 Surface roughness effect:

The results in Figures 9 and 10 show low values for $E$ and $H$ at shallow indentation depths due to the effect of surface roughness as discussed earlier for the bulk samples. There is a generally larger scatter in data at all depths for the films when compared with the bulk samples and this might also be related to the greater roughness of the films. Nevertheless, the standard deviations indicate that for the indentation depth in the range of $0.1$-$0.2t_f$, both the elastic modulus and hardness data were more consistent and reproducible than that for the other depths. This was particularly true for the films sintered at 900 and 1000 °C. For the films sintered at 1100 and 1200 °C, the standard deviations were significantly larger, which can be attributed to their much coarser surface features after sintering as seen in Figures 2 (c) and (d).

#### 3.5.2 Plastic deformation under the indenter:

The effect of densification was shown earlier to have little influence on the elastic modulus measurement for porous bulk samples. However, this is not necessarily the case for films and therefore, further insights into microstructural changes near the indents in the films were obtained using FIB machined cross-sections through the indents. Here a film after sintering at 1000 °C is taken as an example. SEM micrographs of its top surface and cross-section through the middle section of an indent are presented in Fig. 11 (a-c). The spherical tip generates an axisymmetric volume of indented material with its axis at the indent centre and parallel to the direction of indentation. Therefore these images are representative of the whole indented volume. Compressive crushing of the porous structure generated an approximately parabolic "plastic" zone in which the relative density was increased by the crushed debris filling some of the original pore space. Nevertheless, no detectable pile-up or sink-in was observed on the indent surface and no significant cracks were found outside this zone so that the original microstructure was preserved.

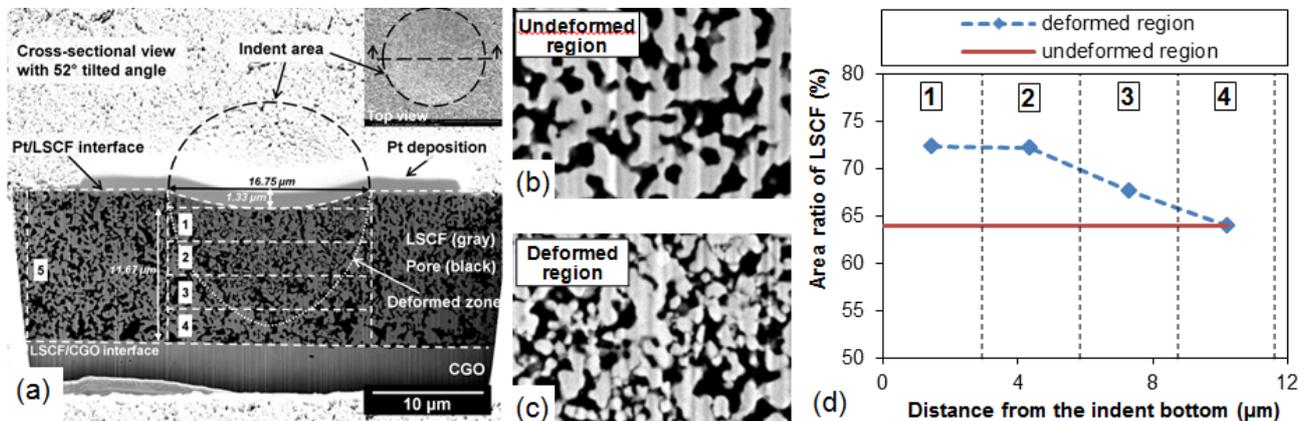

*Fig.11. SEM image of the (a-c) top surface and cross-section of the indent for a LSCF6428 film sintered at 1000 °C; (d) variation of LSCF6428 area ratio with distance from the indent bottom.*

In order to gain a further understanding of the deformation, the deformed area underneath was divided into four rectangular shapes with identical height and width, which was equivalent to the diameter of the projected circular area of the indent. The area ratio of solid material in gray for each rectangle was measured to give the relative "density" of that area using the segmentation module in Avizo software. Fig. 11 (d) shows the variation of this density with distance from the bottom of the indent. In this specimen the density in the undeformed region was measured in area 5 of Fig. 11 (a) to be 63.9%. The density gradually decreased from 72.3% in the area close to the indent, to the average value of 63.9% in the area beyond the influence of the indentation. Thus, in this particular case the "plastic" zone was a crushed region having higher density and extending to a depth of approximately 10 μm.

*3.5.3 Substrate effect:*

In nanoindentation tests of thin films, one of the key difficulties is to avoid errors in measuring film properties caused by influence of the substrate. To do so, an appropriate indentation depth ($h_{max}$) and/or an indent load ($P_{max}$) are desired by taking into account the test film's thickness and the nature of the substrate (whether it has higher or lower elastic modulus than the film). Indenting to a depth less than 10% of the thickness of a film (namely $h_{max}/t_f < 0.1$) has been empirically considered as a safe condition to avoid effects from substrate and extract intrinsic film properties in routine nanoindentation tests[23]. However, it was found in our case that the effect from substrate only became obvious for indentation depth larger than $0.3t_f$, as shown in Figures 9 and 10. Moreover, the influence of the substrate was not as marked for films sintered at 1100 and 1200 °C. This was probably because the stiffness of these films was closer to that of the substrate. Fig. 12 (a) shows the "plastic" zone (higher density caused by crushing deformation) under an indent performed at a relatively deep penetration ($h_{max}/t_f = 0.25$) in a LSCF6428 film after sintering at 1100 °C reaching as far as the interface with the substrate. Measurements of the local "density" using the method described earlier are presented in Fig. 12 (b), which shows that the local density close to the substrate (79.6%) was greater than the density of the material away from the indent (74.4%). This indicates that the substrate had interfered with the progression of the "plastic" zone and, by inference, the elastic zone beyond. Therefore in this particular case the ratio $h_{max}/t_f = 0.25$ was clearly too large for the results not to be influenced by the substrate and, since the substrate possessed a higher modulus (221 GPa) than the film, the measured film modulus $E =$

115 GPa is larger than the true value (which we later deduce to be approximately 90 GPa). To avoid this substrate effect, a lower penetration of around $0.15 t_f$ was applied as shown in Fig. 12 (c). In this case the "plastic" zone had less densification and did not reach the substrate. The modulus of 93 GPa obtained is therefore significantly smaller.

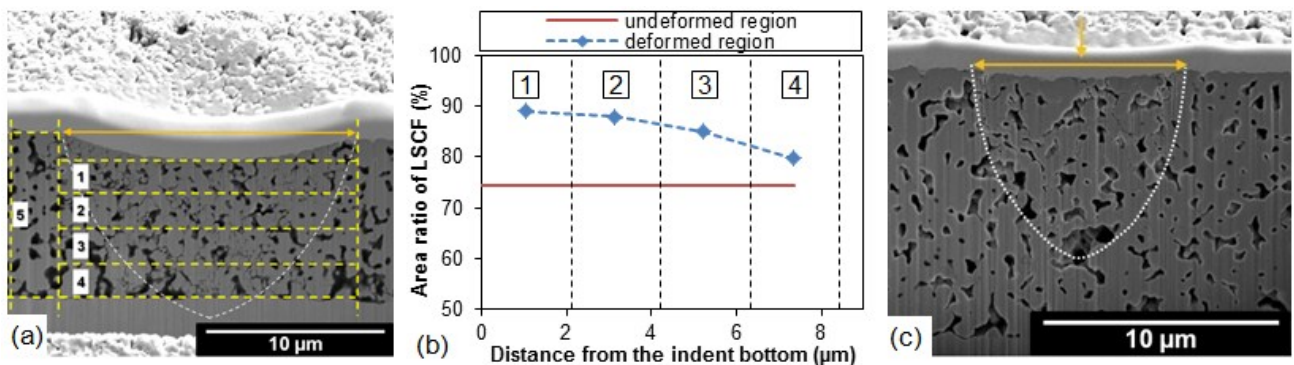

*Fig.12. Densified zone in a film sintered at 1100 °C: (a) cross-section SEM through a relatively deep indent; (b) variation of solid area ratio corresponding to (a); and (c) a shallower indent showing a less pronounced "plastic" zone that did not reach the substrate.*

Fig. 13 shows representative SEM images of the indentation imprints and their cross-sections at indentation depths from 800 to 3600 nm in a film ($t_f \approx 10$ μm) sintered at 1100 °C obtained using the FIB/SEM instrument. The imprints for indentation depth lower than 800 nm (i.e. $h_{max}/t_f$ < 8%) could not be identified with SEM because the residual depths were too shallow. To simplify the image acquisition process, no surface protection coating or impregnation was applied to this sample.

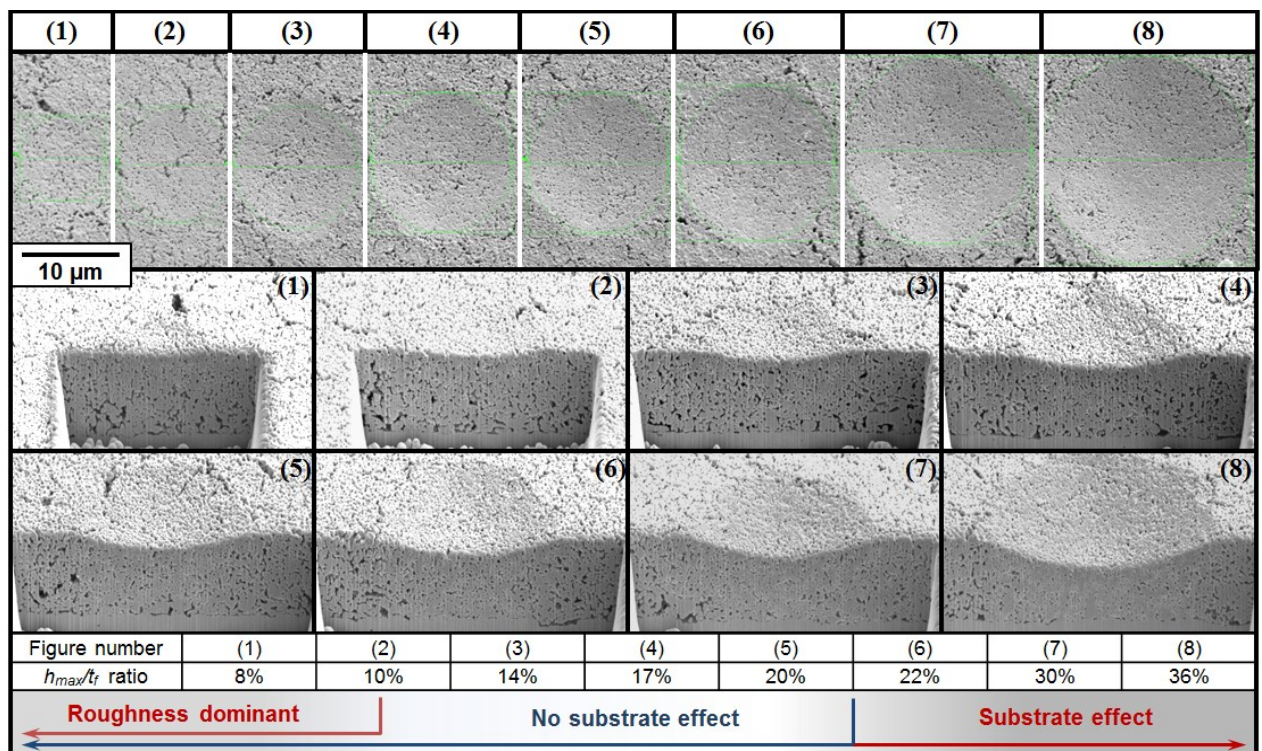

*Fig.13. Top surface and cross-sectional SEM images of indents in a 10 μm thick film corresponding to maximum indentation depths from 800 to 3600 nm (i.e. 8% < $h_{max}/t_f$ < 36%).*

The results clearly show no delamination of the films from the substrates, indicating good interfacial adhesion. Increasing compaction of the film against the substrate can be found at larger indentation depth. In this film a critical indentation depth of approximately 2000 nm ($h_{max}/t_f$ = 20%) can be deduced above which the "plastic" deformation region reached the film/substrate interface and the substrate had a significant effect on the apparent modulus of the film.

Saha and Nix[51] studied the effect of the substrate on nanoindentation for eight different film/substrate systems including metals and ceramics. They concluded that for dense soft film/hard substrate system the substrate hardness had negligible effect on the film hardness because the plastic deformation was always contained within the film and the plastic deformation occurred only when the indenter penetrated the substrate. However, this was not the case in the current study which shows significant impact from the substrate for the film hardness once the indentation depth exceeds 20% of the film thickness. Two significant differences might account for this. First, in the present study a spherical (blunt) indenter was used which resulted in less penetration than a sharp indenter. Second, the films in this study were highly porous and the "plastic" zone, formed by crushing, was denser than the original microstructure of the films. Thus the harder densified material was compressed onto the substrate even when the penetration depth only reached 20% of the film thickness and, for greater indentation depths the substrate had an increasing effect on the measured hardness of the film. In the literature, most studies have adopted the approximate guideline of limiting the indentation depth to less than one tenth of the film thickness in order to obtain reliable values for the elastic modulus of the film[23]. In the present study we have shown that, for the materials investigated here, indentation depths up to 20% of film thickness could be used before the influence of the substrate became much prominent.

### 3.5.4 Elastic modulus and hardness estimations:

It can be concluded from above analysis that when indenting shallower than $0.1t_f$ the results were unduly influenced by surface roughness and above $0.2t_f$ they were increasingly influenced by the substrate. Therefore, the most reliable data points for the films to be extrapolated to estimate true properties in this study are those with the ratio $h_{max}/t_f$ in the range 0.1 to 0.2, where little influence from surface roughness and substrate were introduced. Consequently, extrapolations were carried out to zero $h_{max}/t_f$ using only data from this range, as shown in Figures 9 and 10. The estimated results so obtained are listed in Table 5.

*Table 5. Estimated values of elastic modulus and hardness for LSCF films measured by nanoindentation.*

| Sintering Temperature (°C) | Elastic modulus (GPa) | Hardness (GPa) |
|---|---|---|
| 900 | 32.4±1.2 | 0.37±0.08 |
| 1000 | 48.3±4.6 | 0.61±0.11 |
| 1100 | 90±6.4 | 1.28±0.14 |
| 1200 | 121.5±7.2 | 1.97±0.20 |

As expected, both the elastic modulus and hardness increased dramatically with sintering temperature. For example, the elastic modulus rose nearly fourfold and the hardness rose over sixfold across the whole range of sintering temperatures.

### 3.6 Elastic modulus and hardness vs. porosity for LSCF6428 films and bulk samples

Fig. 14 (a) displays the relationship between the elastic modulus and porosity results obtained using the aforementioned analysis, for both porous films and bulk samples. Note that in this figure the elastic modulus for the 95% dense bulk sample (i.e. 174 GPa) was plotted at zero porosity because as discussed earlier, this was a local measurement of fully dense material (while the IET modulus should be at 5% porosity, which was also shown in the figure). The solid lines in Fig. 14 (a) correspond to exponential relationships between modulus and porosity. Although there could be some other factors controlling the elastic modulus besides porosity, and the relationship of elastic modulus and porosity might not be necessarily exponential, the consistency between the two datasets clearly implies that the porosity played an almost identical key role in both films and bulk samples.

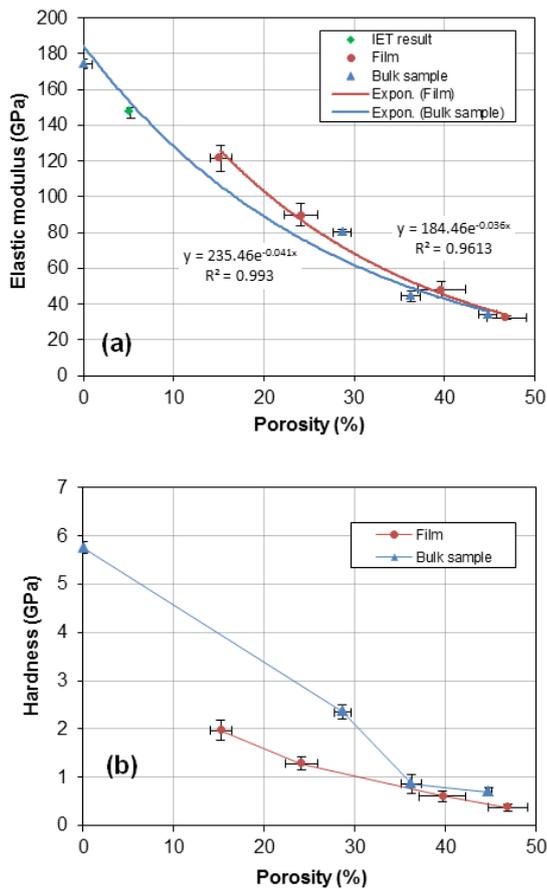

Fig.14. Comparison of (a) elastic modulus and (b) hardness vs. porosity between sintered LSCF films and bulk samples.

It should be noted that for a film sintered on a substrate, the microstructure produced by constrained sintering could differ from that of an as-sintered bulk sample for a given porosity (e.g. by being anisotropic, or having a different ratio of pore size to grain size) and this might account for the small difference between the films and bulk samples. Nevertheless, the effect of porosity on the elastic modulus of both films and bulk can be regarded similar. On the other hand, as shown in Fig. 14 (b), the hardness vs. porosity relationships for those two types of samples require further work due to the complicated plastic deformation mechanism in the porous samples.

### 3.7 Finite element modelling to estimate the elastic modulus of porous films

Fig. 15 shows the 3D models reconstructed from the actual microstructures acquired by FIB/SEM for the porous films after sintering at 900-1200 °C. In the current study, at least 3 representative volumes were sampled for films sintered at each temperature. Poisson's ratio of

0.31[17] and elastic modulus of 174 GPa calculated earlier for the zero-porosity bulk samples were chosen as theoretical elastic parameters of the solid phase in the models. Note that the force applied in the modelling was normal to the film surface. The elastic moduli calculated by FEM are compared with the nanoindentation results in Fig. 16.

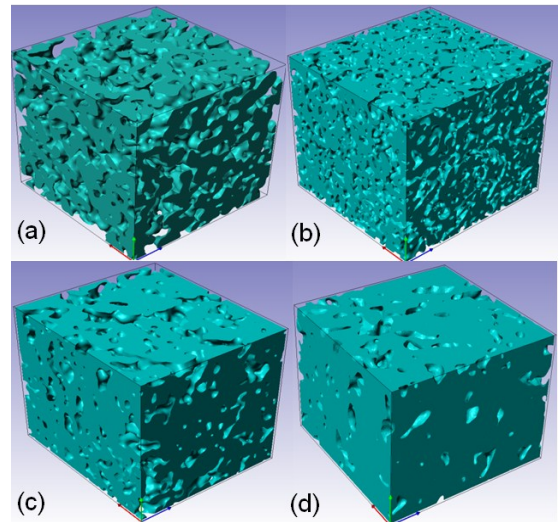

Fig.15. 3D reconstructed models based on the actual 3D microstructural data collected by FIB/SEM of the films after sintering at (a) 900; (b) 1000; (c) 1100; and (d) 1200 °C.

Fig. 16 readily shows that the FEM-derived results agree well with the nanoindentation ones in all cases. The variability of the elastic moduli calculated by FEM is mainly due to the porosity variation in different sampling locations. The FEM result thus validated the method for analyzing the indentation data described earlier.

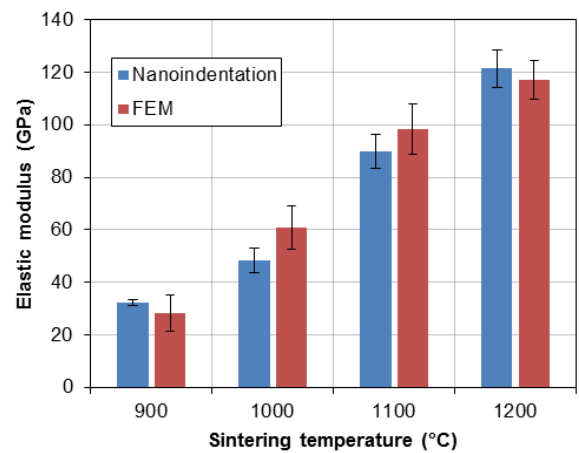

Fig.16. Comparison of elastic modulus measured by nanoindentation and from finite element modelling.

## 4. Conclusions

Bulk specimens of LSCF6428 fabricated by sintering at 900-1200 °C were studied using nanoindentation in order to determine the elastic modulus and hardness of the material. The results were found to be sensitive to surface roughness at shallow indentation depth, while stable values were obtained at larger depths. There was little impact of densification caused by the indentation affecting the measured elastic modulus, but there was a small effect on the hardness. Thus it was found that the elastic modulus of the bulk samples increased from 34 to 174 GPa and hardness from 0.64 to 5.3 GPa with porosity decreasing from 45 to 5% as sintering temperature increased from 900 to 1200 °C. However, the elastic modulus measured by indentation of "dense" bulk specimens sintered at 1200 °C, 174 GPa, was significantly greater than that measured by impulse excitation, 147 GPa. This was shown to be due to residual porosity of approximately 5% in the "dense" specimens which influenced the long range elastic modulus measured by impulse excitation. Therefore the higher value is characteristic of fully dense material. No evidence of a ferroelastic contribution to the load-deflection indentation response was found.

Crack-free films of LSCF6428 of acceptable surface roughness for indentation were prepared by tape casting onto CGO substrates and sintering at temperatures from 900-1200 °C. The porosities of the films were in the range 15-47%. The mechanical properties of the films were investigated using a spherical indenter. The apparent elastic modulus and hardness of the films were found to depend on the ratio of indentation depth to film thickness. They were significantly influenced by surface roughness for shallow indents and by the substrate for deep indents. The influence of surface roughness was due to the granular nature of the porous films, while the influence of the substrate was due to formation of a "plastic" zone of crushed, higher density, material under the indent which touched the substrate if the indentation was too deep. The experiments in this study showed that for this type of porous film and using a spherical indenter, the ratio of indentation depth to film thickness should be kept in the range 0.1 to 0.2, so that reliable film-only values can be obtained by extrapolation to zero load. Thus it was found that the elastic modulus of the films increased from 32 to 121 GPa and hardness from 0.37 to 1.97 GPa as the sintering temperature increased from 900 to 1200 °C and the porosity reduced from 47 to 15%. Comparison with bulk specimens clearly showed that the porous films behaved very similarly to the porous bulk specimens in terms of the dependency of elastic modulus on porosity.

Microstructures obtained by FIB/SEM of the film specimens, after indentation revealed the nature of the "plastic" deformation zone and how this affected the measurement of elastic modulus when it reached the substrate. Finite element analysis was carried out using 3D models reconstructed from the actual microstructures obtained by FIB/SEM tomography. Excellent agreement was found with the results obtained using nanoindentation.

## Acknowledgements

This research was carried out as part of the UK Supergen consortium project on "Fuel Cells: Powering a Greener Future". The Energy Programme is an RCUK cross-council initiative led by EPSRC and contributed to by ESRC, NERC, BBSRC and STFC. Z. Chen is especially grateful to the Chinese Government and Imperial College for financial support in the form of scholarships. Thanks are additionally due to Dr. Farid Tariq for assistance with Avizo and Abaqus software.